| | |
|---|---|
| Title | CO$_2$–CH$_4$ conversion and syngas formation at atmospheric pressure using a multi-electrode dielectric barrier discharge |
| Authors | A. Ozkan[a,b], T. Dufour[a], G. Arnoult[a], P. De Keyzer[a], A. Bogaerts[b], F. Reniers[a] |
| Affiliations | [a] Université Libre de Bruxelles, Chimie analytique et chimie des interfaces, Campus de la Plaine, Bâtiment A, CP255, Boulevard du Triomphe, 1050 Bruxelles, Belgium<br>[b] Universiteit Antwerpen, Research Group PLASMANT, Campus Drie Eiken, Universiteitsplein 1, 2610 Antwerpen-Wilrijk, Belgium |
| Ref. | Journal of CO$_2$ utilization, 2015, Vol. 9, 74-81 |
| DOI | http://dx.doi.org/10.1016/j.jcou.2015.01.002 |
| Abstract | The conversion of CO$_2$ and CH$_4$ into value-added chemicals is studied in a new geometry of a dielectric barrier discharge (DBD) with multi-electrodes, dedicated to the treatment of high gas flow rates. Gas chromatography is used to define the CO$_2$ and CH$_4$ conversion as well as the yields of the products of decomposition (CO, O$_2$ and H$_2$) and of recombination (C$_2$H$_4$, C$_2$H$_6$ and CH$_2$O). The influence of three parameters is investigated on the conversion: the CO$_2$ and CH$_4$ flow rates, the plasma power and the nature of the carrier gas (argon or helium). The energy efficiency of the CO$_2$ conversion is estimated and compared with those of similar atmospheric plasma sources. Our DBD reactor shows a good compromise between a good energy efficiency and the treatment of a large CO$_2$ flow rate. |

# 1. Introduction

Almost 72% of the total greenhouse effect is attributed to water vapor and clouds, the remainder being mainly the result of CO$_2$ [1]. Natural greenhouse gas emissions are responsible for bringing the average temperature of the Earth to +15 °C (instead of –18°C) by absorbing its infrared radiation. However, anthropogenic activities reinforce this situation, leading to an increase of greenhouse gas concentrations in the atmosphere [2,3]. In that respect, carbon dioxide (but also methane) figures among the most important greenhouse gases produced by industries and taking part to the global warming. Its production has increased for many decades. Today, it represents 29 gigatons of emission per year and is expected to increase to 36 or 43 gigatons/year, depending upon the energy world policies, i.e. how we will use existing and new energy sources [4]. For this reason, the remediation of CO$_2$ has received increasing attention in recent years.

Until now, four approaches have been considered to reduce the industrial CO$_2$ footprint: using renewable energy, using non carbon energy resources, CO$_2$ capture and CO$_2$ reforming [5–7]. The latter approach aims at using carbon dioxide as a feedstock and transforming it into value-added products such as carbon monoxide and oxygen, as shown in (1).

$$CO_2 \rightarrow \tfrac{1}{2} O_2 + CO \qquad \Delta G^0_{298K} = +257.2 \text{ kJ.mol}^{-1} \quad (1)$$

This aforementioned reaction is thermodynamically limited and highly endothermic. According to Le Chatelier's principle, a high reaction temperature and a low CO$_2$ partial pressure are required to achieve a high conversion [7–9]. Owing to the high thermodynamic stability of the CO$_2$ molecule in standard conditions, its dissociation can only be achieved through endothermic reactions requiring an external energy source. In that respect, conventional chemistry processes have already been used, such as electroreduction of CO$_2$ [6]. Besides, non-thermal atmospheric plasma processes can be employed such as corona discharges [10,11], dielectric barrier discharges (DBD) [12–18], gliding-arcs [19,20] and plasma jets [21,22]. Low pressure plasma sources can also be used such as microwave discharges [23,24]. Among these sources, most of the energy required for the dissociation of $CO_2$ depends on the electron energy distribution function (EEDF). Carbon dioxide can be mixed with methane to form carbon monoxide and molecular hydrogen in (2), but also other products of interest can be formed, such as oxygenated organic molecules and hydrocarbons [25,26].

$$CO_2 + CH_4 \rightarrow 2H_2 + 2CO \qquad \Delta G^0_{298K} = 170.8 \text{ kJ.mol}^{-1} \qquad (2)$$

The conversion of $CO_2$ and $CH_4$ by an atmospheric dielectric barrier discharge (DBD) is reported in this study, using Ar as a carrier gas to generate more metastable species and therefore stabilize the discharge. Using a tubular DBD offers a promising and innovative solution since the transformation of $CO_2$ can be performed ''on line'', i.e. directly at the output of industrial chimneys instead of releasing the $CO_2$ into the atmosphere and hence increase the greenhouse effect. Therefore, it does not require capture, transport or storage of $CO_2$ and, for instance, could partially close the carbon loop if coupled to green electricity. By using gas chromatography (GC), we demonstrate that this process is efficient to obtain CO and value-added products. Three parameters are evaluated: the $CO_2$ and $CH_4$ flow rates, the power supplied to the DBD and the nature of the carrier gas (Ar or He). The energy efficiency of the $CO_2$ conversion is estimated and compared with those of similar plasma sources.

## 2. Experimental set-up

### 2.1. DBD reactor

A cylindrical multi-electrode DBD reactor dedicated to the treatment of elevated gas flow rates has been designed as shown in Fig. 1. It consists of a 2 mm thick tube made in quartz with an external diameter of 34 mm and a length of 100 mm (so as to ensure a long residence time). The gas enters via 16 inlets of 0.75 mm in diameter arranged into a circular pattern, then travels longitudinally through the tubular reactor and finally flows out of the reactor via 16 outlets (same configuration as the inlet). The discharge is generated between six AC high-voltage tubular electrodes set at equal distance from a central tubular electrode which is grounded. The power applied to the high-voltage electrodes is provided by an AFS Generator G10S-V with a maximum power of 1000 W and a variable frequency in the range between 1 and 30 kHz. The distance between the grounded electrode and each high-voltage electrode is the same as the distance between two high-voltage electrodes, namely 3 mm. The grounded electrode is a copper rod with a diameter of 5 mm and a length of 100 mm, while the high-voltage electrodes are copper wires approximately 1 mm in diameter and with the same length of 100 mm. The high-voltage electrodes are encompassed into alumina dielectric tubes with 0.75 mm thickness, as depicted in Fig. 1.

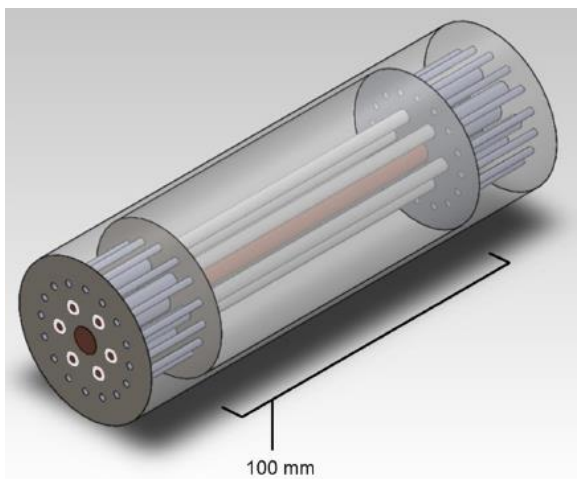
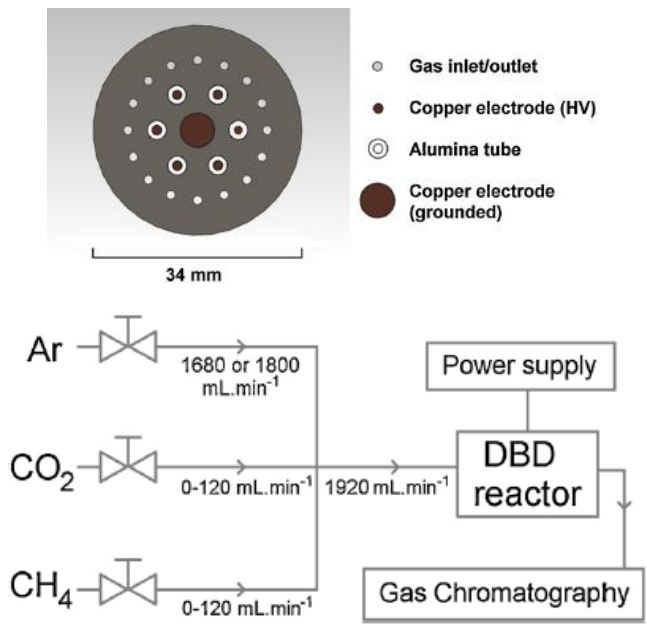

Fig. 1. Schematic diagram of the DBD reactor.

Fig. 2. Schematic diagram of the entire experimental set up.

## 2.2. Entire set-up

A schematic of the entire experimental setup is shown in Fig. 2. Argon, carbon dioxide and methane are introduced into the reactor via Aalborg volumetric flow meters able to measure flow rates as high as 1800, 120 and 120 mL.min$^{-1}$, respectively. Argon (or helium) is used as the carrier gas to initiate and maintain the discharge. The total flow rate of the gas mixture supplying the DBD reactor remains fixed at 1920 mL.min$^{-1}$ while the $CO_2$ and $CH_4$ flow rates are both varied from 0 to 120 mL min$^{-1}$.

The products resulting from the plasma phase reactions are analyzed downstream of the reactor with an online gas chromatograph (Agilent 6890N) equipped with a 60/80 Carboxen 1000 column (Supelco 1-2390-U). The products are analyzed with two detectors: a thermal conductivity detector (TCD) and a flame ionization detector (FID). The conversion of $CO_2$ and $CH_4$ are calculated according to Eqs. (3) and (4), respectively, where A represents the peak area assigned to $CO_2$ or $CH_4$ in the chromatogram:

$$CO_{2\,Conversion}\,(\%) = \chi_{CO_2} = \frac{A_{CO_2\,without\,plasma} - A_{CO_2\,with\,plasma}}{A_{CO_2\,without\,plasma}} \times 100 \qquad (3)$$

$$CH_{4\,Conversion}\,(\%) = \chi_{CH_4} = \frac{A_{CH_4\,without\,plasma} - A_{CH_4\,with\,plasma}}{A_{CH_4\,without\,plasma}} \times 100 \qquad (4)$$

The selectivities of $H_2$, $O_2$, CO, $C_2H_6$ and $C_2H_4$ have been calculated as reported in Table 1, listed as H, O or C based selectivities, depending on the plasma composition ($CH_4$, $CO_2$, $CO_2/CH_4$ respectively).

# 3. Results

## 3.1. Effect of the $CO_2$ and $CH_4$ flow rates

The plasma is generated in a mixture of $CO_2$, $CH_4$ and Ar (or He) to investigate the effect of the reactive gas flow rates on their conversion. The Ar flow rate is set to 1800 mL.min$^{-1}$ while the $CO_2$ and $CH_4$ flow rates can be tuned between 0 and 120 mL min$^{-1}$, but the sum of both is always equal to 120 mL min$^{-1}$. Fig. 3 represents the $CO_2$ and $CH_4$ conversions as a function of the $CO_2$ and $CH_4$ flow rates. Both for $CO_2$ and $CH_4$, an increase in the flow rate is always correlated with a decrease in its conversion. Indeed, for $CO_2$ flow rates increasing from 20 to 120 mL.min$^{-1}$, $\chi_{CO2}$ decreases from 8.3% to 6.1% while $\chi_{CH4}$ decreases from 21.5% to 10.9% when the $CH_4$ flow rates rise from 20 to 120 mL.min$^{-1}$. This figure illustrates also that $CH_4$ is always converted to a larger extent than $CO_2$, whatever the individual gas flow rates. Chemical reactions in the plasma lead to the dissociation of these molecules, thus generating products that can also recombine to form new species such as $H_2$, $O_2$, CO, $C_2H_4$ and $C_2H_6$ whose volumetric fractions ($f_V$) are plotted in Fig. 4(a) as a function of the $CO_2$ and $CH_4$ flow rates. Each $f_V$ fraction is calculated as the ratio of the product flow rate to the $CO_2/CH_4$ mixture flow rate, multiplied by 100. The main products are molecular hydrogen ($f_{V,max}[H_2]$= 7.73%), carbon monoxide ($f_{V,max}[CO]$= 8.13%) and molecular oxygen ($f_{V,max}[O_2]$= 3.98%), the latter being detected only if no $CH_4$ is injected in the discharge. Other products such as ethylene and ethane are also formed but in smaller proportions ($f_{V,max}[C_2H_4]$= 0.52% and $f_{V,max}[C_2H_6]$= 1.51%). The production of CO is more important with an increase in the $CO_2$ flow rate, reaching a plateau of approximately 8.10% for $CO_2$ flow rates higher than 80 mL.min$^{-1}$. In the same way, the production of hydrogen, ethane and ethylene increases with the $CH_4$ flow rate. The production of $O_2$ is only present for pure $CO_2$ plasma while it disappears after $CH_4$ addition. That probably means that the CH4 reactive species interact with oxygen in the discharge. It is quite logical that the decomposition of $CO_2$ favors the production of CO and $O_2$ while the decomposition of CH4 leads to the production of $H_2$, $C_2H_4$ and $C_2H_6$ but also of carbon black powder (not detected by gas chromatography).

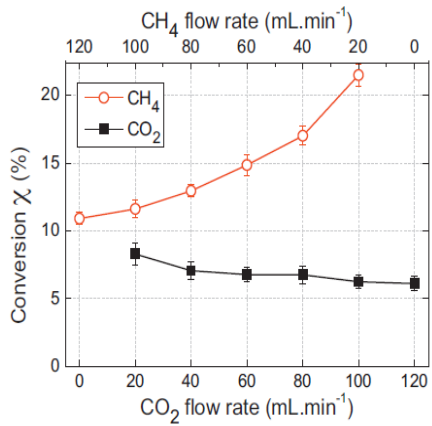

Fig. 3. $CO_2$ and $CH_4$ conversions as a function of the $CO_2$ and $CH_4$ flow rates with $\Phi_{Tot}$ = 1920 mL.min$^{-1}$, $\Phi_{Ar}$ = 1800 mL.min$^{-1}$, $\Phi_{CO2}$ = $\Phi_{CH4}$ = 120 mL min$^{-1}$, plasma power = 45 W, frequency = 19.5 kHz.

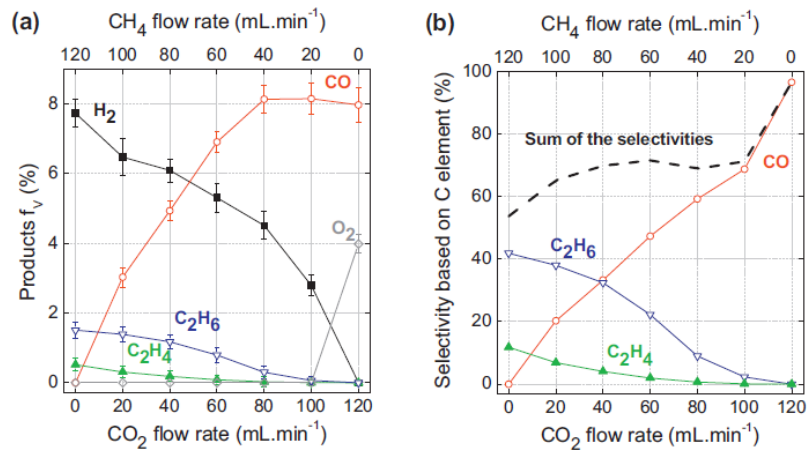

Fig. 4. (a) Volumetric fractions of $H_2$, CO, $O_2$, $C_2H_4$ and $C_2H_6$ (using the TCD) and (b) C-based selectivity of the quantified gaseous products as a function of the $CO_2$ and $CH_4$ flow rates with $\Phi_{Tot}$ = 1920 mL.min$^{-1}$, $\Phi_{Ar}$ = 1800 mL.min$^{-1}$, $\Phi_{CO2}$ = $\Phi_{CH4}$ = 120 mL min$^{-1}$, plasma power = 45 W, frequency = 19.5 kHz.

The selectivities of these products have also been calculated using the formulas from Table 1 and considering three cases:

(1) When using CH4 as unique reactive gas, the H-based selectivities are approximately 26% for $H_2$, 32% for $C_2H_6$ and 6% for $C_2H_4$. As $S_H(H_2) + S_H(C_2H_6) + S_H(C_2H_4) + S_H(other) = 100\%$, other products may be considered, i.e. $CH_x$ non-gaseous products which are assumed to deposit on the inner walls of the reactor, especially on the surface of the high voltage central electrode.

(2) When using $CO_2$ as unique reactive gas, the O-based selectivities lead to $S_O(O_2) + S_O(CO) = 48.2 + 49.2 = 97.4$. This value is very close to 100% and if experimental errors are considered – in particular the uncertainties of the flowmeters (<3%) – then we can conclude that $CO_2$ is virtually only converted to molecular oxygen and carbon monoxide.

(3) When using a $CH_4/CO_2$ mixture, the C-based selectivities of three gaseous carbonated products (namely CO, $C_2H_6$ and $C_2H_4$) are plotted as a function of $CO_2$ flow rate in Fig. 4(b). According to the relation $S_C(CO) + S_C(C_2H_6) + S_C(C_2H_4) + S_C(other) = 100\%$, an increase in $S_C(other)$ is evidenced with a rise in the $CH_4$ flow rate. We believe that this can be due to solid carbon deposit, formation of acetylene and liquid/gaseous formaldehyde.

| | H-based selectivities ($CH_4$ plasma) | O-based selectivities ($CO_2$ plasma) | C-based selectivities ($CH_4/CO_2$ plasma) |
|---|---|---|---|
| $S_{H_2}$ (%) | $\dfrac{n_{H_2}^{produced}}{2 \cdot n_{CH_4}^{converted}} \times 100$ | – | – |
| $S_{O_2}$ (%) | – | $\dfrac{n_{O_2}^{produced}}{n_{CO_2}^{converted}} \times 100$ | – |
| $S_{CO}$ (%) | – | $\dfrac{n_{CO}^{produced}}{2 \cdot n_{CO_2}^{converted}} \times 100$ | $\dfrac{n_{CO}^{produced}}{n_{CO_2}^{converted} + n_{CH_4}^{converted}} \times 100$ |
| $S_{C_2H_4}$ (%) | $\dfrac{n_{C_2H_4}^{produced}}{n_{CH_4}^{converted}} \times 100$ | – | $\dfrac{2 \cdot n_{C_2H_4}^{produced}}{n_{CO_2}^{converted} + n_{CH_4}^{converted}} \times 100$ |
| $S_{C_2H_6}$ (%) | $\dfrac{3 \cdot n_{C_2H_6}^{produced}}{2 \cdot n_{CH_4}^{converted}} \times 100$ | – | $\dfrac{2 \cdot n_{C_2H_6}^{produced}}{n_{CO_2}^{converted} + n_{CH_4}^{converted}} \times 100$ |

Table 1 Formulas for the H, O or C based selectivities of $H_2$, $O_2$, CO, $C_2H_6$ and $C_2H_4$ (n is the number of moles).

The case of solid carbon deposit on the central copper electrode area clearly appears after a few minutes of plasma treatment. The apparent granular texture of this deposit may be responsible for local electrical peak effects, thus leading to a more filamentary discharge. As a result, the $CO_2$ conversion would change in case of prolonged use of the reactor. No arc has been formed, which otherwise would have prematurely deteriorated the barrier, and hence the durability of the reactor. To prevent these problems, coke deposit can easily be removed by cleaning the inner walls of the reactor and polishing them with sandpaper. Another convenient way is to apply a pure $CO_2$ or pure $O_2$ plasma to remove the coke deposit.

## 3.2. Effect of the power

Fig. 5 shows the $CH_4$ and $CO_2$ conversions versus the power applied to the DBD in the range between 30 W and 80 W for $\Phi_{Ar}$ = 1680 mL min⁻¹, $\Phi_{CO2} = \Phi_{CH4}$ = 120 mL.min⁻¹ and an AC frequency of 19.5 kHz. The $CO_2$ conversion increases from 2.0% to 7.5% upon rising power, while the $CH_4$ conversion increases from 6.7% to 14.8% in the same power range. The two conversions can be considered as linearly increasing with the power since their correlation coefficients are $r^2(CO_2) = 0.976$ and $r^2(CH_4) = 0.899$. The slopes of both curves are almost the same, consistently with the results of Zheng et al. performed in a two-electrode DBD reactor [27]. It is also clear that the methane conversion is always higher than $\chi_{CO2}$ (difference of at least 5%) thanks to its lower bond energy. The volumetric fractions of $H_2$, CO, $C_2H_4$, $C_2H_6$ plotted in Fig. 6 versus the power indicate that the production of syngas (hydrogen and carbon monoxide) also increases linearly with the power, and both products are formed nearly equally, yielding a syngas ratio close to 1. A linear increase is also observed in the case of $C_2H_4$ and $C_2H_6$, although the slopes are less significant.

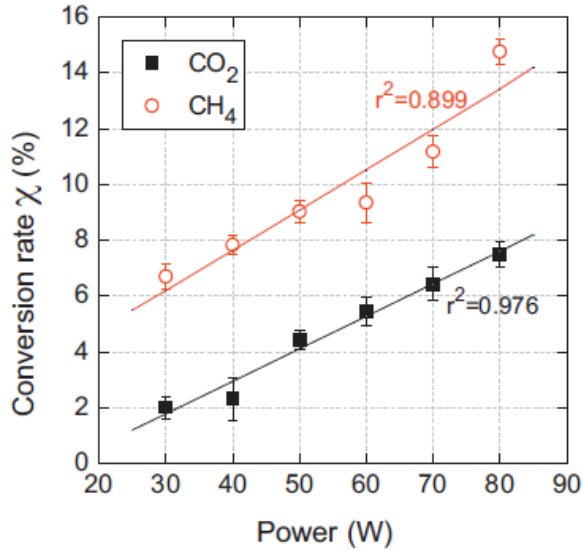
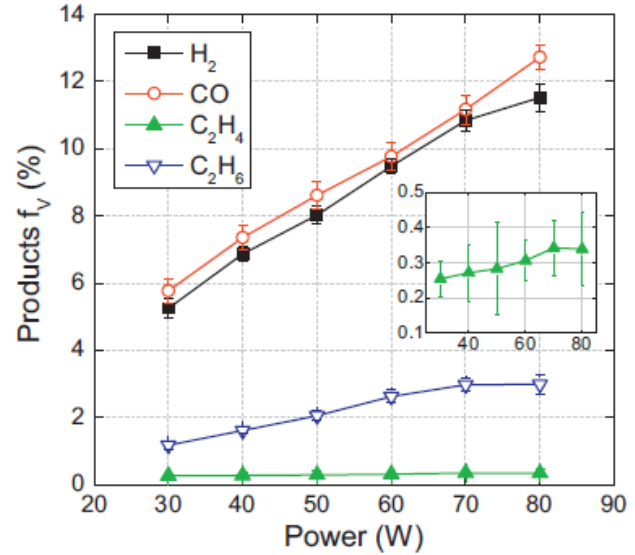

Fig. 5. Conversions of $CO_2$ and $CH_4$ versus the power ($\Phi_{Ar}$ = 1680 mL.min$^{-1}$; $\Phi_{CO2}$ = $\Phi_{CH4}$ = 120 mL.min$^{-1}$; frequency = 19.5 kHz).

Fig. 6. Volumetric fractions of $H_2$, CO, $C_2H_4$ and $C_2H_6$ (using the TCD) versus the power ($\Phi_{Ar}$ = 1680 mL.min$^{-1}$; $\Phi_{CO2}$ = $\Phi_{CH4}$ = 120 mL min$^{-1}$; frequency = 19.5 kHz).

### 3.3. Effect of the carrier gas

The influence of the carrier gas (argon or helium) is investigated for the same flow rate set to 1800 mL.min$^{-1}$ and the reactive gas flow rates set to $\Phi_{CO2}$ = $\Phi_{CH4}$ = 60 mL.min$^{-1}$. The nature of the carrier gas seems to have an important impact on the conversion of $CO_2$ and $CH_4$; see Fig. 7a. The conversion of $CH_4$ is indeed higher in the presence of helium than with argon (respectively 21.4% and 16.4%) while the opposite effect is observed for the conversion of $CO_2$ since $\chi_{CO2}$ = 6.8% with helium and $\chi_{CH4}$ = 11.5% with argon. It is also worth mentioning that for the same plasma power (60 W) and frequency (17.1 kHz), a filamentary discharge and a glow discharge are obtained with argon and helium, respectively, as shown in Fig. 7b and c.

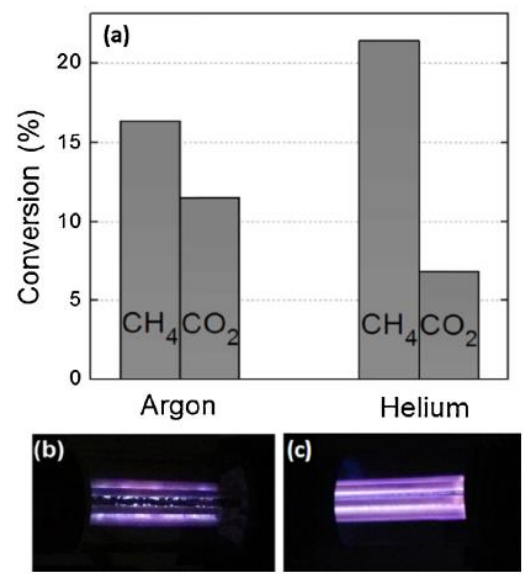

Fig. 7. (a) Conversions of $CH_4$ and $CO_2$ in Ar/$CO_2$/$CH_4$ and He/$CO_2$/$CH_4$ plasmas with $\Phi_{Tot}$ = 1920 mL.min$^{-1}$, $\Phi_{Ar}$ or $\Phi_{He}$ = 1800 mL.min$^{-1}$, $\Phi_{CO2}$ = $\Phi_{CH4}$ = 60 mL.min$^{-1}$, plasma power = 60 W and frequency = 17.1 kHz; (b) picture of the Ar/$CO_2$/$CH_4$ discharge, illustrating the filamentary behavior; (c) picture of the He/$CO_2$/$CH_4$ discharge, illustrating the glow mode.

# 4. Discussion

## 4.1. Effect of the $CO_2$ and $CH_4$ flow rates on the plasma reactivity and reaction products formed in the $CO_2/CH_4$ conversion process

### 4.1.1. Overview of the important reactions

Plasmas are complex media where several hundred reactions of production and consumption can occur [28,29]. The most plausible mechanisms for the formation and consumption of intermediate and value-added products in the $CO_2/CH_4$ gas mixture are listed in Table 2. In the following sections, we explain how the most important reaction products are formed.

*Table 2 Reaction pathways for the formation and consumption of intermediate and value-added products of $CH_4$ and $CO_2$ conversion.*

| | Reaction | Rate constant | Ref. |
|---|---|---|---|
| (R1) | $CH_4 + e \rightarrow CH_3^\bullet + H^\bullet + e$ | $\sigma_3^*$ | [30] |
| (R2) | $CH_2 + O^\bullet \rightarrow CO + H_2$ | $5.53 \times 10^{-11}$ cm$^3$ s$^{-1}$ | [31] |
| (R3) | $CH_3^+ \rightarrow CH_2^\bullet + H^\bullet$ | $1.69 \times 10^{-08}$ cm$^3$ s$^{-1}$ | [32] |
| (R4) | $CH_3^+ + e \rightarrow CH_2^\bullet + H^\bullet$ | $2.25 \times 10^{-08}$ cm$^3$ s$^{-1}$ | [30,33] |
| (R5) | $CH_3^+ + e \rightarrow CH^\bullet + H_2$ | $7.88 \times 10^{-09}$ cm$^3$ s$^{-1}$ | [30,33] |
| (R6) | $CH_3^\bullet + H^\bullet \rightarrow CH_2^\bullet + H_2$ | $1.00 \times 10^{-10}$ cm$^3$ s$^{-1}$ | [32] |
| (R7) | $CH_4 + H^\bullet \rightarrow CH_3^\bullet + H_2$ | $5.83 \times 10^{-13}$ cm$^3$ s$^{-1}$ | [32] |
| (R8) | $CH_4^+ + H^\bullet \rightarrow CH_3^+ + H_2$ | $1.00 \times 10^{-11}$ cm$^3$ s$^{-1}$ | [34] |
| (R9) | $H^\bullet + H^\bullet \rightarrow H_2$ | $1.44 \times 10^{-14}$ cm$^3$ s$^{-1}$ | [35] |
| (R10) | $CO_2 + H^\bullet \rightarrow CO + OH^\bullet$ | $1.40 \times 10^{-29}$ cm$^3$ s$^{-1}$ | [36] |
| (R11) | $CH^\bullet + O^\bullet \rightarrow CO + H^\bullet$ | $6.90 \times 10^{-11}$ cm$^3$ s$^{-1}$ | [32] |
| (R12) | $CO_2 + CH^\bullet \rightarrow 2CO + H^\bullet$ | $9.68 \times 10^{-13}$ cm$^3$ s$^{-1}$ | [31] |
| (R13) | $CH_3^\bullet + CH_3^\bullet \rightarrow C_2H_6$ | $4.20 \times 10^{-11}$ cm$^3$ s$^{-1}$ | [37] |
| (R14) | $C_2H_5^\bullet + H^\bullet \rightarrow C_2H_6$ | $2.25 \times 10^{-10}$ cm$^3$ s$^{-1}$ | [38] |
| (R15) | $C_2H_5^\bullet + CH_4 \rightarrow C_2H_6 + CH_3^\bullet$ | $1.83 \times 10^{-24}$ cm$^3$ s$^{-1}$ | [36] |
| (R16) | $C_2H_6 + e \rightarrow C_2H_5^\bullet + H^\bullet + e$ | $\sigma_{18}^*$ | [39] |
| (R17) | $C_2H_5^\bullet + e \rightarrow C_2H_4 + H^\bullet + e$ | $\sigma_{19}^*$ | [39] |
| (R18) | $CH_4 + CH^\bullet \rightarrow C_2H_4 + H^\bullet$ | $9.74 \times 10^{-11}$ cm$^3$ s$^{-1}$ | [32] |
| (R19) | $CH_3^\bullet + CH_2^\bullet \rightarrow C_2H_4 + H^\bullet$ | $7.01 \times 10^{-11}$ cm$^3$ s$^{-1}$ | [32] |
| (R20) | $C_2H_5^\bullet + O^\bullet \rightarrow C_2H_4 + OH^\bullet$ | $4.40 \times 10^{-11}$ cm$^3$ s$^{-1}$ | [31] |
| (R21) | $CO_2 + e \rightarrow CO + O^\bullet + e$ | $\sigma_{23}^*$ | [40] |
| (R22) | $CO_2 + e^- \rightarrow CH_4CO + O^\bullet + e^-$ | $\sigma_{24}^*$ | [40] |
| (R23) | $CO_2 + O^\bullet \rightarrow CO + O_2$ | $2.01 \times 10^{-10}$ cm$^3$ s$^{-1}$ | [41] |
| (R24) | $CO_2^+ + e \rightarrow CO + O^\bullet$ | $2.71 \times 10^{-07}$ cm$^3$ s$^{-1}$ | [33] |
| (R25) | $CH_3^\bullet + O^\bullet \rightarrow H_2CO + H^\bullet$ | $1.12 \times 10^{-10}$ cm$^3$ s$^{-1}$ | [42] |
| (R26) | $CH_2^\bullet + O_2 \rightarrow H_2CO + O^\bullet$ | $5.39 \times 10^{-13}$ cm$^3$ s$^{-1}$ | [32,43] |

### 4.1.2. Production of hydrogen

Several chemical reactions give rise to the production of molecular hydrogen through the dissociation of hydrocarbon species upon collision with an atom (R2), an electron (R5) or an H radical (R6, R7 and R8). The rate constants of these reactions are in the order of $10^{-13}$–$10^{-10}$ cm$^3$.s$^{-1}$, except for R5 which is somewhat higher ($7.88*10^{-9}$ cm$^3$.s$^{-1}$) as the collision occurs between an energetic electron and an ion. Although the rate constant of R9 is a bit lower than the other ones ($1.44*10^{-14}$ cm$^3$.s$^{-1}$), the recombination of two H radicals may be considered as very important since $v = k_{11}[H]^2$ and H is produced in many other reactions such as R3, R4, R11, R12, R16, R17, R18, R19 and R25. Electron impact reactions R16 and R17 are not described with a rate constant but with a cross section s which depends on the electron temperature.

### 4.1.3. Production of CO

The formation of CO is directly correlated with the dissociation of $CO_2$. The reactions responsible for the production of CO are given by R2, R10, R11, R12, R21, R22, and R23 [44]. (R21) is electron impact dissociation of $CO_2$ into CO and O, which is the most important process in $CO_2$ splitting. When $CH_4$ is present, the O atoms will be further consumed by R2, R11, R20, R23 and R25 and this explains the higher $CO_2$ conversion when more $CH_4$ is present in the gas mixture. Indeed, as stated by the Le Chatelier's principle, the dissociation is more favorable as one (or both) of the reaction products is constantly consumed. This effect has been demonstrated in the literature: Tagawa et al. have observed an increasing $CO_2$ conversion by placing an $O_2$ trapper membrane into a $CO_2/CH_4$ discharge in order to separate $O_2$ from the gas stream. As a consequence, the $CO/CO_2$ equilibrium is more shifted to CO [45].

#### 4.1.4. Production of ethane

The recombination of two $CH_3^\bullet$ radicals can lead to the production of ethane according to reaction (R13). R14 and R15 could also lead to the production of ethane but are less probable. Indeed, as computed by Snoeckx et al. in the case of a similar atmospheric DBD source supplied in $CH_4$–$CO_2$, the density of $CH_3$ is always higher than the one of $C_2H_5$ [29].

#### 4.1.5. Production of ethylene

The formation of ethylene may result from a two-step collisional mechanism, where first an electron collision leads to the dissociation of $C_2H_6$ into $C_2H_5^\bullet$ and H radicals (R16), followed by a second electron collision with $C_2H_5^\bullet$ resulting in the abstraction of a H radical to produce ethylene (R17). This simple mechanism can explain why $f_V[C_2H_4]$ is always lower than $f_V[C_2H_6]$.

#### 4.1.6. Other reaction products

Formaldehyde traces have also been detected. Their formation can result from $CH_3$ radicals (R25) or to a lower extent from $CH_2$ radicals (R26). Other oxygenated products have not been detected at the conditions under study, probably because their amounts are under the limit of detection of the gas chromatography detectors. According to the literature, the formation of other oxygenated organic molecules such as acetic acid or methanol may also occur in a plasma [17,46,47].

The higher volumetric fraction of $H_2$, compared to ethane and ethylene, can be explained according to several chemical reactions (R2, R5, R6, R7, R8 and R9). Indeed, there are more reactions for $H_2$ and $H^\bullet$ formation compared to reactions for $C_2H_6$ and $C_2H_4$ formation. Moreover, there are more reactions consuming $C_2H_6$ or $C_2H_4$ than consuming $H_2$. $C_2H_6$ or $C_2H_4$ is indeed very easily consumed once it is produced. That is why the $H_2$ amount is always higher than the amounts of $C_2H_6$ and $C_2H_4$.

## 4.2. Effect of the power

The linear increase of $CO_2$ and $CH_4$ conversions as a function of the power results from a linear increase in the electron density (Fig. 5). Indeed, the dissociation of C–H and C=O bonds requires energies of a few eV that may be mostly transferred from the electrons. An increase in the plasma power can induce higher electron temperatures and higher electron densities. In our case, the increase in electron temperature may be assumed as negligible since in a classical DBD, it would induce a stronger filamentary regime that has not been observed here. Increasing the plasma power can also induce higher electron densities that can be assumed as linearly depending on the power if the electron permeability and the electric field profile are considered as weakly dependent on the applied power.

For the production of $C_2H_4$ and $C_2H_6$, a linear increase upon increasing power is also observed, but the slopes are less pronounced than for CO and $H_2$. This is probably due to the fact that the production of these molecules is not simply based on one electron impact reaction, like the formation of $H_2$ from $CH_4$ and the splitting of $CO_2$ into CO. Indeed, in order to obtain $C_2H_6$, two $CH_3^\bullet$ radicals are necessary (R13) while to obtain $C_2H_4$, two electronic collisions with $C_2H_6$ are required (R16 and R17).

## 4.3. Effect of the carrier gas

According to Fig. 7, $\chi_{CH4}$ is always higher than $\chi_{CO2}$ whatever the nature of the carrier gas. Indeed, in a plasma, the dissociation of $CH_4$ is easier than for $CO_2$ since the bond dissociation energy of C–H (4.48 eV) is lower than the bond dissociation energy of C=O (5.52 eV) [48].

However, the fact that $CH_4$ is more efficiently dissociated in He than in Ar, whereas $CO_2$ is more efficiently dissociated in Ar than in He, is less straightforward. The reason is that the shape of the electron energy distribution function (EEDF) is different when the plasma is in the filamentary regime (Ar) or in the glow regime (He). The EEDF of these two regimes is sketched in Fig. 8, assuming Maxwellian distributions (thermodynamic equilibrium) for the sake of clarity [49]. The bond dissociation energies of C–H and C=O are also reported in Fig. 8. In the filamentary regime, the EEDF is characterized by (i) a number of warm electrons much lower than in a glow discharge but also by (ii) a tail extending toward higher energies, meaning that the hot electrons (even if not in a large number) can be involved into new collisional processes, which require a stronger activation energy [50]. In the case of the $CH_4$ dissociation, all the electrons that contribute to breaking of the C–H bonds, must be located at the right side of BDE(C–H) and under the EEDF curves: this corresponds to the area $A_1$ in the glow regime (He) and $A_3$ in the filamentary regime (Ar) (see insert in Fig. 8). As $A_1$ is larger than $A_3$, more electrons can participate to the dissociation of $CH_4$ in the case of He, hence this explains why $(\chi_{CH4})_{He} > (\chi_{CH4})_{Ar}$.

On the other hand, a higher electron energy is needed for breaking the C=O bonds of $CO_2$: all electrons that contribute to this bond breaking, must be located at the right side of BDE(C=O), and under the EEDF curves: this corresponds to area $A_2$ in the glow regime (He) and to area $A_4$ in the filamentary regime (Ar). As $A_4$ is larger than $A_2$, more electrons can participate to the dissociation of $CO_2$ in Ar than in He, and this explains why $(\chi_{CO2})_{Ar} > (\chi_{CO2})_{He}$. Finally, if we consider the areas which correspond to the electrons that can contribute to the dissociation of C–H and C=O bonds for both the glow and filamentary regimes, it appears that $A_1 > A_3 > A_4 > A_2$. Hence, this corresponds to $(\chi_{CH4})_{He} > (\chi_{CH4})_{Ar} > (\chi_{CO2})_{Ar} > (\chi_{CO2})_{He}$, which is indeed observed in Fig. 7. Therefore, the $CO_2$ conversion is the lowest in helium since $A_2$ is the smallest among the four areas. In other terms, the number of electrons available in a He discharge for the conversion of $CO_2$ is very small as the energy of these electrons has to be equal to or higher than the activation energy to break C=O (i.e. 5.52 eV). In summary, the nature of the carrier gas – and consequently the regime (glow or filamentary) of the DBD – directly impacts the shape of the EEDF and therefore the electron collision processes that may occur.

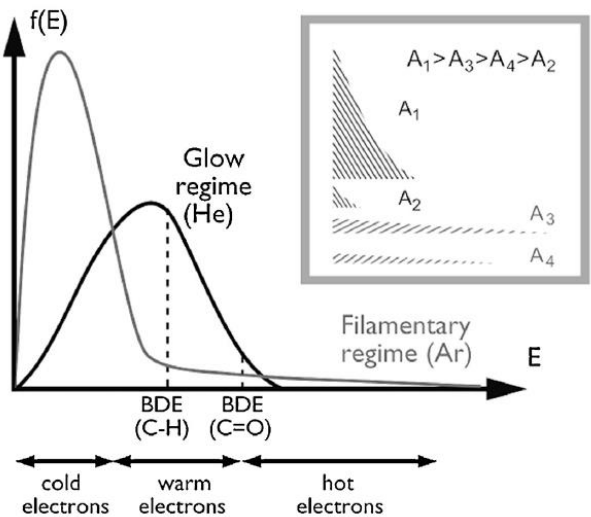

*Fig. 8. Schematic sketch of the EEDFs in the case of a glow discharge (He) and a filamentary discharge (Ar) at thermodynamic equilibrium. Also indicated are the bond dissociation energies (BDE) for C–H and C=O bonds. The insert shows the fractions of electrons that can contribute to dissociation of C–H and C=O bonds in both regimes (see text for more explanation).*

### 4.4. Conversion, specific energy input and energy efficiency: comparison with literature

The specific energy input (SEI) corresponds to the energy density ($E_d$) in J.cm$^{-3}$, and can also be expressed in eV molecule$^{-1}$ as defined by the following equations:

$$E_d \, (J \, cm^{-3}) = \frac{Power \, (J \, s^{-1})}{Gas \, flow \, rate \, (cm^3 \, s^{-1})} \quad (5)$$

$$SEI \, (eV \, molecule^{-1}) = \frac{E_d \, (J \, cm^{-3}) \times 6.24 \times 10^{18} \, (eV \, J^{-1}) \times 24500 \, (cm^3 \, mol^{-1})}{6.022 \times 10^{23} \, (molecule \, mol^{-1})} \quad (6)$$

The energy efficiency of the CO$_2$ conversion ($\eta_{CO2}$) has been calculated (in %) from the conversion $\chi_{CO2}$, the enthalpy of (2) namely $\Delta H^0_{298K}$ = 247.3 kJ.mol$^{-1}$ = 2.56 eV molecule$^{-1}$ and the SEI value, according to the following equation:

$$\eta_{CO_2} \, (\%) = \frac{\chi_{CO_2} \, (\%) \times \Delta H^0_{298 \, K} \, (ev \, molecule^{-1})}{SEI \, (eV \, molecule^{-1})} \quad (7)$$

The same equation can be written for the energy efficiency of the CH$_4$ conversion ($\eta_{CH4}$). Hence, the energy efficiency is separately defined for CO$_2$ and CH$_4$ in this article. Eq. (7) indicates that an increase in the SEI systematically induces a decrease in $\eta$, at least when the conversion stays constant, and this means that we should have a SEI value as low as possible to obtain a more energy efficient process. This is indeed clear from Fig. 9, where the energy efficiencies of both CH$_4$ and CO$_2$ clearly drop upon higher SEI. For a SEI as low as 5.7 eV molecule$^{-1}$, $\eta_{max}$(CO$_2$) = 3.3% while $\eta_{max}$(CH$_4$) = 4.9%.

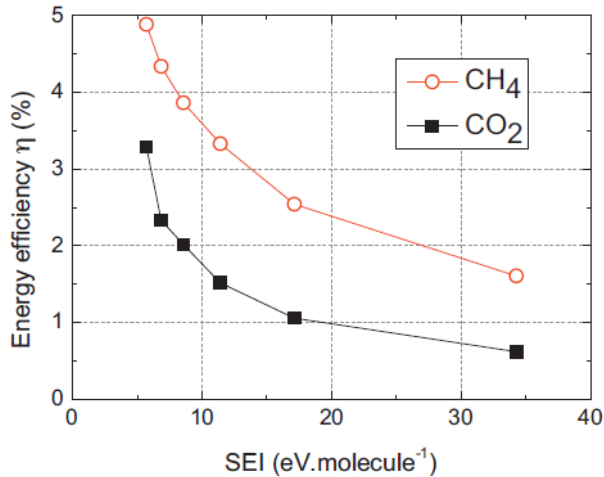

*Fig. 9. Energy efficiency as a function of specific energy input in our experimental set-up with $\Phi_{Tot}$ = 1920 mL.min$^{-1}$, $\Phi_{Ar}$ = 1800 mL.min$^{-1}$, $\Phi_{CO2}$ = $\Phi_{CH4}$ = 120 mL.min$^{-1}$, plasma power = 45 W, frequency = 19.5 kHz.*

A comparison of our multiple electrode DBD reactor with other atmospheric plasma sources is presented in Table 3. This table reports various plasma sources: DBD, AC glow discharges, pulsed corona and gliding arcs with different geometries and specific operating parameters, namely: frequency, power given by authors, nature of the carrier gas or reactive gas, and CO$_2$ flow rate. Note that some of these experiments apply to pure CO$_2$ splitting, while others refer to dry reforming (i.e., conversion of both CO$_2$ and CH$_4$). However, we focus here only on the CO$_2$ conversion. Also, it should be noted that some experiments were carried out for the pure greenhouse gases, while others made use of a carrier gas. The conversion and energy efficiency are in general higher in a carrier gas but it is obviously less interesting for applications. The optimal CO$_2$ conversions for all these cases are plotted in Fig. 10 as a function of the corresponding SEI while their energy efficiencies are plotted in Fig. 11 as a function of the CO$_2$ flow rates. In these figures, each squared number refers to one of the plasma sources listed in Table 3.

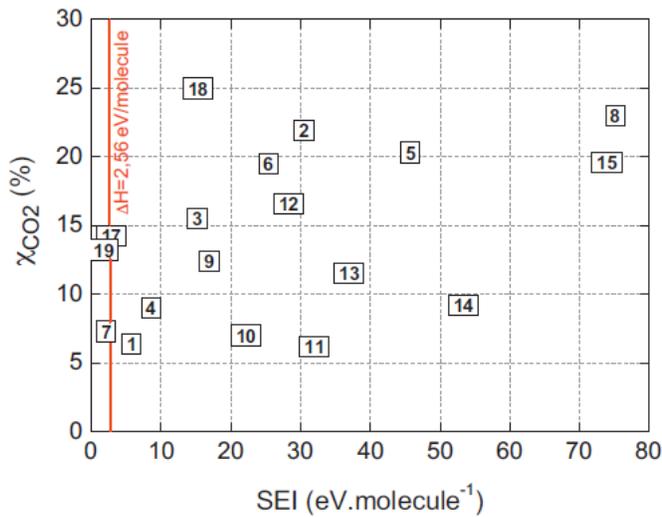

Fig. 10. CO$_2$ conversions vs SEI for the various plasma sources listed in Table 3.

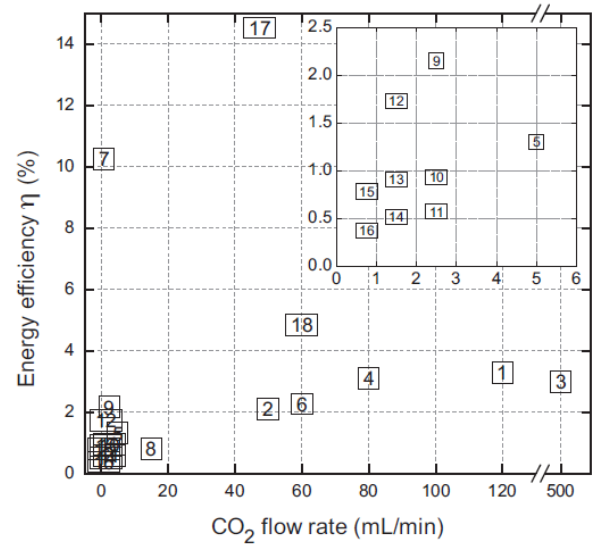

Fig. 11. Energy efficiency (%) of the various atmospheric plasma sources listed in Table 3, as a function of the CO$_2$ flow rates. The area in the left bottom is enlarged as an insert, placed in the right top of the figure.

A first remark is that no atmospheric plasma source can reach a $\chi_{CO2}$ higher than 25%. Moreover, no general trend can be deduced: the cloud of points indicates that some plasma sources are very energy-consuming with high $\chi_{CO2}$ (#8, #14 and #15) while some others are much more dedicated to CO$_2$ reforming at a lower energy cost (#1, #4, #7, #17, #18 and #19) since they are located close to the vertical line at 2.56 eV molecule$^{-1}$, standing for the enthalpy of reaction (2). The plasma source #7 shows a good energy efficiency, but is not suitable as it can handle CO$_2$ flow rates of only 0.8 mL min$^{-1}$. The plasma sources #17 and #19 present interesting conversions for SEI as low as ours, but with the disadvantage of their geometry, which is a pulsed corona and a gliding arc, respectively. Indeed, the advantage of using a tubular DBD lies in the ability to place it at the nozzle exit of a combustion process to treat the entire gas flow since all the gas passes through the discharge zone. On the contrary, a pulsed corona and a gliding arc can exhibit ''dead volumes'' where the gas passes through the reactor without being treated in the plasma zone. Furthermore, the corona source is not adapted for high flow rates treatment as its discharge volume is not that important, which makes it a good candidate only to handle low flow rates. The plasma source #4 is an interesting alternative to our plasma process. In our case, the CO$_2$ conversion is not so high but the SEI is quite low, so this yields a good energy efficiency, as shown in Fig. 11 (#1). Our plasma source shows a good compromise between a high energy efficiency and the treatment of a significant CO$_2$ flow rate, probably thanks to the multielectrode configuration.

| # | Plasma source | | Frequency (kHz) | Power (W) | Carrier or reactive gas | $CO_2$ flow rate (mL min$^{-1}$) | Ref. |
|---|---|---|---|---|---|---|---|
| | Type | Geometry | | | | | |
| 1 | DBD | Tube multi-electrodes | 19.5 | 45 | Ar | 120 | Our reactor |
| 2 | DBD | Tube | 30 | 100 | / | 50 | [16] |
| 3 | DBD | Plane | 30 | 500 | $CH_4$ | 500 | [47] |
| 4 | DBD | Tube | 2.2 | 45 | Ar | 80 | [27] |
| 5 | DBD | Plane | 25 | 15 | He/$CH_4$ | 5 | [26] |
| 6 | DBD | Tube | 25 | 100 | $CH_4$ | 60 | [51] |
| 7 | DBD | Tube | 8.1 | 0.11 | / | 0.8 | [44] |
| 8 | DBD | Tube | 20 | 74 | $CH_4$ | 15 | [52] |
| 9 | AC glow discharge | Tube | 8.1 | 2.78 | He | 2.5 | [53] |
| 10 | | | 8.1 | 3.64 | Ar | 2.5 | |
| 11 | | | 8.1 | 5.25 | $N_2$ | 2.5 | |
| 12 | | | 8.1 | 2.78 | He | 1.5 | |
| 13 | | | 8.1 | 3.64 | Ar | 1.5 | |
| 14 | | | 8.1 | 5.25 | $N_2$ | 1.5 | |
| 15 | | | 8.1 | 3.64 | Ar | 0.75 | |
| 16 | | | 8.1 | 5.25 | $N_2$ | 0.75 | |
| 17 | Pulsed corona | Electrode tip | 20–200 | 9 | / | 47.5 | [54] |
| 18 | DBD | Tube | 30 | 60 | Ar/$CH_4$ | 60 | [55] |
| 19 | Gliding arc | "V" shaped electrode | 20 | 225 | / | 2000 | [20] |

Table 3 Comparison of various plasma sources dedicated to the conversion of $CO_2$ at atmospheric pressure. The $CO_2$ conversions vs SEI and the energy efficiency as a function of $CO_2$ flow rate for all these cases are reported in Figs. 10 and 11, respectively.

# 5. Conclusion

The production of syngas (CO and $H_2$), $C_2H_4$ and $C_2H_6$ has been achieved at atmospheric pressure in a dielectric barrier discharge operating in $CO_2$ and $CH_4$, with Ar or He as carrier gases. The main mechanisms responsible for the production of these compounds have been discussed. In this study, the effect of the concentration of $CO_2$/$CH_4$ in the mixture on the conversion has been demonstrated. Furthermore, the effect of power has also been reported, showing a linear increase in the $CO_2$ and $CH_4$ conversions but also in the production of syngas as a function of the supplied power. Finally, the energy efficiency of the $CO_2$ conversion has been calculated and compared with those of other atmospheric plasma sources. Our DBD reactor offers very encouraging results as it offers one of the best compromises between a high energy efficiency and the treatment of a large $CO_2$ flow rate.

# 6. Acknowledgement

The authors acknowledge financial support from the IAP/7 (Inter-university Attraction Pole) program 'PSI-Physical Chemistry of Plasma-Surface Interactions', financially supported by the Belgian Federal Office for Science Policy (BELSPO).